\documentclass[final,5p,times,twocolumn,11pt]{elsarticle}

\usepackage{amssymb,amsmath,delarray,graphicx,multicol,esint,hyperref}
\usepackage{lineno}
\journal{Physics Letters A}

\begin{document}

\begin{frontmatter}

\title{Convection driven by internal heating}

\author[David]{David Goluskin\corref{cor}}
\address[David]{Department of Applied Physics and Applied Mathematics, Columbia University, \\New York, NY 10027, USA}
\cortext[cor]{Corresponding author}
\ead{dg2422@columbia.edu}
\author[Ed]{Edward A. Spiegel}
\address[Ed]{Department of Astronomy, Columbia University, New York, NY 10027, USA}

\begin{abstract}
Two-dimensional direct numerical simulations are conducted for convection sustained by uniform internal heating in a horizontal fluid layer. Top and bottom boundary temperatures are fixed and equal. Prandtl numbers range from 0.01 to 100, and Rayleigh numbers ($R$) are up to $5\cdot10^5$ times the critical $R$ at the onset of convection. The asymmetry between upward and downward heat fluxes is non-monotonic in $R$. In a broad high-$R$ regime, dimensionless mean temperature scales as $R^{-1/5}$. We discuss the scaling of mean temperature and heat-flux-asymmetry, which we argue are better diagnostic quantities than the conventionally used top and bottom Nusselt numbers.
\end{abstract}


\end{frontmatter}

\section{Introduction}

Sustained, thermally driven convection occurs when heat is continually injected into a fluid, causing some of the fluid to be hotter and less dense than the fluid directly above it. This heat injection may be accomplished through the thermal boundary conditions, through volumetric heat sources, or both. Standard Rayleigh-B\'enard (RB) convection in a layer of fluid is driven by keeping the bottom boundary fixed at a higher temperature than the top. This is the most studied convection scenario, having become a favored model for investigating instabilities, bifurcations, pattern formation, and thermal turbulence \cite{Getling1998, Siggia1994}. Convection driven by internal heat sources has been studied much less, though it plays a fundamental role in several geophysical, astrophysical, and industrial processes. Here we study the extreme case of convection driven solely by internal heating with no heat injected at the boundaries.

Two simple configurations stand out in the literature on internally heated layers: one that is insulated below with the top boundary temperature fixed, and the other with the top and bottom boundary temperatures fixed and equal to one another. We present 2D numerical findings for the latter case, which is the more challenging one due to the presence of a stably stratified bottom boundary layer. We focus on both the qualitative nature of the flow and on certain significant integral quantities at large rates of internal heating (that is, at high Rayleigh numbers). The 1985 review of Kulacki and Richards \cite{Kulacki1985} discusses theoretical, experimental, and numerical efforts on both configurations, and the 1987 review of Cheung and Chawla \cite{Cheung1987} contains some general discussion of internally heated convection, though it deals primarily with the insulating-bottom case. We mention below some more recent works of relevance to our present equal-boundary-temperature configuration.

In several physical experiments during the nineteen-seventies \cite{Kulacki1972, Kulacki1974, Jahn1974}, convection was studied in fluid layers heated internally by electric currents and with the top and bottom boundary temperatures kept as close to equal as possible. A more recent experiment used periodically distributed heaters in air \cite{Lee2007}, and 2D \cite{Jahn1974} and 3D \cite{Worner1997} simulations have also been performed. Some related configurations have also been simulated to study accident scenarios in nuclear reactor engineering \cite{Nourgaliev1997, Grotzbach1999, Horvat2001, Liu2006, Chen2009}. However, such studies often employ geometries and boundary conditions that are motivated by particular applications and from which it is hard to draw conclusions about the basic plane layer problem. They also often resort to turbulence models to simulate the highly turbulent flows that can occur in a nuclear melt. Herein, we return to the plane layer setup and attack it with direct numerical simulations.

Geophysics offers natural occurrences of convection driven by a combination of internal heating and boundary effects. In the Earth's mantle, convection is sustained both by radiogenic heating throughout the mantle itself and by conduction from the underlying hot outer core \cite{Sotin1999, Travis1994, Houseman1988}. In the upper atmosphere, convection is driven by radiative cooling throughout and by heating from the lower atmosphere and the surface of the earth \cite{Berlengiero2012}. (With a change of variables, this is identical to fluid heated internally and cooled from above.) We anticipate that further results on internally driven convection will supplement the abundant prior results on boundary-driven convection in helping to understand the essential features of such dually driven flows.

Astrophysics provides instances of internally driven convection that is both compressible and nonuniformly heated. In the cores of stars on the main sequence, heat is produced by thermonuclear reactions. In the most massive of such stars, the heating rate is very sensitive to temperature, and this sensitivity creates steep thermal gradients that drive powerful convection in the cores \cite{Kippenhahn1994}. Clearly, this and many other instances of internally driven convection contain more complications than we confront here. Nonetheless, our findings on uniform heating allow comparison with laboratory experiments, and they may help elucidate the physics of the more involved applications.

The diverse instances of internally heated convection display a great range of Prandtl numbers, from the extremely low effective Prandtl numbers of astrophysical plasmas to the essentially infinite values in the mantle. Prandtl numbers at the lower end of this range are prohibitively expensive to simulate at large Rayleigh numbers, but we have made some effort to study the role of Prandtl number by simulating values between 0.01 and 100. Values near the bottom of this range arise in reactor engineering \cite{Arcidiacono2001}, and the top of this range is, by most measures, near the infinite-Prandtl number limit that mantle studies invariably adopt.

The next section introduces the model to be studied. Section \ref{sec: qual} describes the qualitative results of our 2D simulations. In section \ref{sec: int} we discuss key integral quantities, while in section \ref{sec: quant} we present our simulation results on these quantities, along with phenomenological scaling arguments. Section \ref{sec: con} concludes the letter.

\section{Governing equations}
\label{sec: eq}

Our nondimensionalization is typical for internally heated convection and goes back at least to Roberts \cite{Roberts1967}. As is standard in the study of RB convection \cite{Chandrasekhar1981}, we nondimensionalize length by the domain height, $d$, and time by the characteristic thermal time, $d^2/\kappa$, where $\kappa$ is thermal diffusivity.  The dimensionless spatial domain is thus bounded horizontally by $0\le x \le A$ and vertically by $-1/2\le z\le1/2$, where $A$ is the aspect ratio. In boundary-driven convective flows, such as RB convection, the boundary conditions provide a temperature scale. Our boundary conditions provide no such scale, so we instead make use of $H$, the product of the volumetric heating rate and the heat capacity. We nondimensionalize temperature by $d^2H/\kappa$, which is the increase in temperature that an insulated parcel of fluid would undergo in one unit of conductive time. The dimensionless equations of motion in the Boussinesq approximation\footnote{Strictly speaking, this is not the standard Boussinesq approximation given in \cite{Chandrasekhar1981} because the symmetry $(z,T)\mapsto(-z,-T)$ is absent.}
are then
\begin{align}
\nabla \cdot \mathbf u &= 0 \label{eq: inc} \\
\partial_t \mathbf u + \mathbf u \cdot \nabla \mathbf u &=
	-\nabla p + \sigma \nabla^2 \mathbf u + \sigma R T\mathbf{\hat z} \label{eq: u} \\
\partial_t T + \mathbf u \cdot \nabla T &= \nabla^2 T + 1. \label{eq: T}
\end{align}
The two dimensionless parameters are a Rayleigh number, $R$, that is a variant of the standard one, and the standard Prandtl number, $\sigma$,
\begin{align}
R &= \frac{g\alpha d^5H}{\kappa^2\nu} \label{eq: R} \\
\sigma &= \frac{\nu}{\kappa},
\end{align}
where $\nu$ is kinematic viscosity, $g$ is gravitational acceleration in the $-\mathbf{\hat z}$ direction, and $\alpha$ is the linear coefficient of thermal expansion.

\subsection{Boundary conditions}

We impose no-slip and fixed-temperature conditions at the top and bottom boundaries,
\begin{equation}
\mathbf u = T = 0 \text{ at } z=\pm\tfrac{1}{2},
\end{equation}
where the governing equations' invariance under a uniform shift in temperature allows us to choose boundary temperatures of zero for convenience. These same boundary conditions have been employed in several experiments \cite{Kulacki1985} and in the variational computation of a lower bound on the mean temperature \cite{Lu2004}. The equality of the top and bottom boundary temperatures ensures that the volume-averaged vertical heat transport by conduction is zero at all times, though, as we shall see, convection transports heat upward on average.

Our simulations employ periodic side boundaries. However, the integral relations of section \ref{sec: int} below also hold for boundaries that are stress-free and insulating (that is, $\mathbf u \cdot \mathbf n$, $\mathbf n \cdot \nabla \mathbf u$, and $\mathbf n \cdot \nabla T$ vanish, where $\mathbf n$ is the outward unit normal to the side boundaries), as well as on 3D domains with analogous boundary conditions. The choice of side boundaries should not affect mean quantities in the infinite-aspect-ratio limit that we strive to approximate in our simulations.

\subsection{Conductive solution}

The simplest solution to the governing equations is the conductive solution, which is defined by motionless fluid and a horizontally uniform, vertically parabolic temperature profile that we denote by $\widetilde T$,
\begin{equation}
\widetilde T(z) = \tfrac{1}{2}\left(\tfrac{1}{4}-z^2\right).
\end{equation}
In the conductive state, heat flows outward across the top and bottom boundaries at equal rates. Analyzing $\widetilde T$ by standard energy stability methods \cite{Joseph1976}, we find that $R<26,927$ suffices to guarantee that $\widetilde T$ is the unique stable solution on a horizontally infinite domain, and any perturbations decay exponentially. Otherwise, the conductive solution is unstable to infinitesimal perturbations of horizontal wavenumber 4.00 when $R>R_L=37,325$ and the domain admits the wavenumber \cite{Sparrow1963, Kulacki1975a}. This value of $R_L$ is corroborated by our simulations of (\ref{eq: inc})-(\ref{eq: T}). Transient growth likely occurs between $R_E$ and $R_L$. We have not observed sustained convection in simulations at any subcritical Rayleigh numbers, but this possibility could be better explored by performing amplitude expansions near $R_L$.

\section{Qualitative results}
\label{sec: qual}

The equations of motion were simulated in 2D using the nek5000 spectral element code \cite{nek} on computational domains wide enough to approximate certain bulk properties of a horizontally infinite domain. Details on convergence criteria for meshes, time-averages, and aspect ratios appear in \ref{sec: comp}. Visualizations were created using VisIt \cite{Childs2011}.

\subsection{Steady rolls}

\begin{figure}
\centering
\includegraphics[scale=.65]{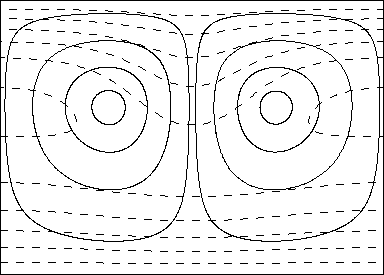}
\caption{Streamlines (solid) and isotherms (dashed) for a pair of stable steady rolls at $R=50,000$ and $\sigma=1$. The aspect ratio of $A=1.4$ is approximately equal to the wavelength that arises naturally in a large 2D domain. The left-hand roll rotates clockwise, while its mirror image rotates counterclockwise. The temperature changes by 0.02 between isotherms, increasing from the zero at the boundaries to 0.12 on the innermost isotherm.}
\label{fig: steady}
\end{figure}

Once $R$ exceeds $R_L$, the critical value for linear instability, pairs of steady rolls form, as illustrated in Fig.\ \ref{fig: steady}. These rolls lack the up-down symmetry of RB rolls. This asymmetry has been noted in previous studies, such as \cite{Kulacki1972}, and its origin is clear: the unstable temperature gradient near the upper boundary drives the flow, while stably stratified fluid near the lower boundary inhibits it. As a result, fluid flows across the top and downward faster than it flows across the bottom and upward. Conservation of mass thus dictates that the down-flow regions are narrower than the up-flow regions, and that the roll centers lie above the midline. The cold top boundary layer thickens in the down-flow regions to form nascent thermal plumes, while the cold bottom boundary layer is much more horizontally uniform. Correspondingly, at larger $R$, we shall see that numerous well-defined plumes descend from the top boundary layer, while cold fluid leaves the bottom boundary layer only when it is stirred up by the interior flow. Such up-down asymmetry is typical of penetrative convection (as in \cite{Zahn1991}, for instance).

\subsection{Near-periodicity in time}
\label{sec: per}

As the Rayleigh number is raised, the steady roll pairs become increasingly asymmetric until they lose stability. In 2D, time is the only dimension available for the rolls to break symmetry, and indeed the flow begins to oscillate in time for large enough $R$. At moderate $\sigma$ and large $A$, steady rolls are replaced by oscillatory ones well before $R$ reaches $2R_L$, in contrast to 2D RB flow, which remains steady for $R$ hundreds of times larger than $R_L$. Correspondingly, time-dependence was observed in internally heated experiments for $R$ not much larger than $R_L$ \cite{Kulacki1972}.

\begin{figure}
\centering
\includegraphics[scale=1]{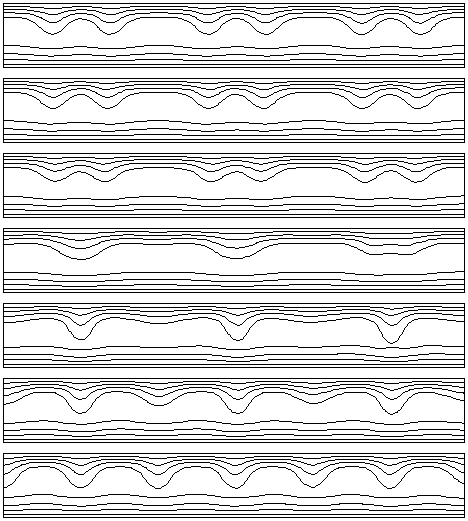}
\caption{(Color video online; see \ref{sec: videos}) Temperature isotherms for a series of time slices depicting the fast phase of an oscillation at $R=65,000$ and $\sigma=1$, cropped to an aspect ratio of 8 from simulations with $A=12$. The temperature changes by $0.025$ between isotherms, increasing from zero at the boundaries to 0.1 on the two central isotherms. The merging and genesis of cold thermal plumes is evident, going forward in time from top to bottom with time steps of 0.15. By contrast, the slow phases of the oscillations have periods on the order of 100.}
\label{fig: fast}
\end{figure}

Every oscillating solution we have observed is of the same type -- highly nonlinear relaxation oscillations that are nearly periodic in both time and in the horizontal, but never exactly so. At the start of the slow phase, the cold plumes are spaced nearly uniformly. The spacing becomes less uniform as each plume becomes a member of a pair, drifting toward its mate at a gradually accelerating rate. When the two plumes are sufficiently close together, the flow enters the fast phase, which is depicted in Fig.\ \ref{fig: fast}. In this phase, plumes quickly collide and merge. But even as the number of plumes is being halved, new plumes are already forming in each gap, restoring the original number and restarting the slow phase. Qualitatively similar merging and genesis of plumes has been seen in 3D simulations \cite{Worner1997} and laboratory experiments \cite{Kulacki1972}, though the nearly periodic and spatially synchronized relaxation oscillations we observe have not been reported previously.

The deviation from exact periodicity may or may not be due to numerical inaccuracies. It is possible that the underlying solutions are in fact stable, exactly periodic relaxation oscillations, but that numerical noise always significantly perturbs the oscillations when time derivatives are very small during the slow phase. Alternately, we may be observing chaotic oscillations. These possibilities invite a more detailed study of the system's bifurcation structure.

\subsection{Toward turbulence}
\label{sec: turb}
The Rayleigh number ranges over which spatially synchronized oscillations have been observed are all rather narrow, the change in $R$ over such ranges being always much smaller than $R_L$. When $R$ is above these ranges, plumes continue to grow from the top boundary layer and merge with others, but these events are no longer synchronized across the spatial domain, and the flow is no longer nearly periodic in time. The flow becomes visibly more irregular as the Rayleigh number is raised further, but because the flow is constrained to 2D, solutions always exhibit some roll-like coherence. In 3D, on the other hand, rolls will quickly lose stability to three-dimensional structures \cite{Kulacki1972}. Nonetheless, we regard the 2D system as interesting in its own right, and for moderate-to-high $\sigma$, the 2D version is apparently a good predictor of the 3D system's integral quantities \cite{Schmalzl2004}.

\begin{figure}
\centering
\includegraphics[scale=.42]{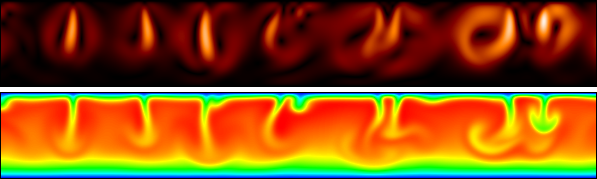}
\caption{Instantaneous fluid speed (top) and temperature (bottom) at $R=10^7$ and $\sigma=5$. The faster-moving fluid is lighter in the speed field. The hottest fluid is red in the temperature field, and the coldest is blue.}
\label{fig: Ra 1E7}
\end{figure}

\begin{figure}
\centering
\includegraphics[scale=.42]{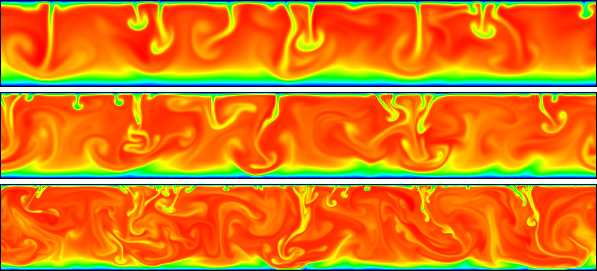}
\caption{(Videos online for $R$ of $10^8$ and $10^{10}$; see \ref{sec: videos}) Instantaneous temperature fields for Rayleigh numbers of $10^8$ (top), $10^9$, and $10^{10}$ with $\sigma=5$. The hottest fluid in each image is red, and the coldest is blue, though the color scales are normalized differently in each image.}
\label{fig: plumes}
\end{figure}

As $R$ is increased into the millions, flows at moderate-to-high $\sigma$ exhibit mushroom-like cold plumes descending from the top boundary layer. (Such structures are familiar from the study of RB convection, along with their hot counterparts that rise from the bottom boundary layer. See \cite{Zocchi1990}, for example.) For $\sigma=5$, Fig.\ \ref{fig: Ra 1E7} shows typical fluid speed and temperature fields at $R=10^7$, and Fig.\ \ref{fig: plumes} shows temperature fields up to $R=10^{10}$, by which point we are beginning to see the eddies and filaments characteristic of 2D turbulence. (These and subsequent visualizations have aspect ratios of 7 but are cropped from simulations of larger $A$.) In the fluid speed and temperature fields of Fig.\ \ref{fig: Ra 1E7}, strong down-flow evidently aligns with thermal plumes, while weaker recirculation occurs in between plumes. Though not shown, circulation is similarly aligned with clusters of plumes in the higher-$R$ cases of Fig.\ \ref{fig: plumes}.

The thermal plumes change in several ways as $R$ is increased. Individual plumes become smaller because they scale with the upper thermal boundary layer, which thins as $R$ is raised. Meanwhile, they also become more numerous, and they show an increasing tendency to merge with nearby plumes. At large $R$, most individual plumes merge with others to become part of larger composite plumes, as in the $R=10^9$ and $R=10^{10}$ fields of Fig.\ \ref{fig: plumes}. Unlike the small individual plumes, these composite plumes are able to penetrate to the bottom thermal boundary layer, driving roll-like structures whose heights and widths are comparable to the height of the domain. The scale of the composite plumes is maintained because the number of their constituent plumes increases as the constituent plumes themselves shrink. (In 3D RB simulations, plumes similarly cluster to form larger structures whose horizontal scale varies only weakly with the Rayleigh number \cite{VonHardenberg2008}.) The increasingly strong composite plumes also drive increasingly powerful and disordered interior flow, which at large $R$ begins to stir up cold fluid from the bottom boundary layer. In Fig.\ \ref{fig: plumes}, cold ejections from the bottom boundary layer are evident at $R=10^9$ and quite pronounced at $R=10^{10}$.

\subsection{Low Prandtl numbers}

\begin{figure}
\centering 
\includegraphics[scale=.42]{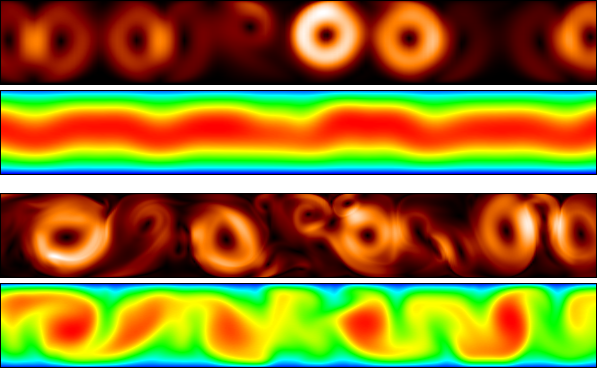}
\caption{(Videos online; see \ref{sec: videos}) Instantaneous fluid speeds (above) and temperatures (below) at $\sigma=0.01$ with $R=2\cdot10^5$ (top pair) and $R=10^7$ (bottom pair). The faster-moving fluid is lighter in the speed fields. The hottest fluid is red in the temperature fields, and the coldest is blue.}
\label{fig: low Pr} 
\end{figure}

Flows of low Prandtl number differ significantly from those of moderate-to-high Prandtl number, most notably in the decrease of up-down asymmetry at a given $R$ and in the appearance of so-called flywheels, which have been studied in low-$\sigma$ RB convection \cite{Moore1973, Proctor1977, Clever1981}. Fig.\ \ref{fig: low Pr} shows typical low-$\sigma$ fluid speed and temperature fields at two Rayleigh numbers. In the $R=2\cdot10^5$ speed field, two flywheels are evident in the right half of the domain. Each flywheel was part of a pair of rolls until it subsumed its mate and became more axisymmetric.  Eventually, such a flywheel loses momentum and becomes part of a pair again, but new ones come into existence repeatedly. In the $R=10^7$ field, flywheels dominate the flow, and the hottest fluid is found solely in their centers, rather than being distributed throughout the interior, as it is in the $\sigma=5$ field for the same $R$ that is shown in Fig.\ \ref{fig: Ra 1E7}. We simulated Prandtl numbers only as low as $0.01$ since computation becomes increasingly expensive as $\sigma$ is lowered beyond unity. This is because the thinning kinetic boundary layers require finer spatial meshes, and the slower advective dynamics require longer dimensionless times for spatiotemporal averages to converge. To access very small $\sigma$ at large $R$, one might instead simulate the small-$\sigma$ limit of the Boussinesq equations \cite{Spiegel1962, Thual1992}.

\section{Integral quantities}
\label{sec: int}

Let an overbar denote an average over the horizontal and time, and let angle brackets denote an average over the entire domain and time, as in
\begin{align}
\overline f(z) &:= \lim_{\tau\to\infty}\frac{1}{\tau}\int_0^\tau dt \frac{1}{A}\int_0^{A} dx~f(x,z,t) \label{eq: area} \\
\langle f \rangle &:= \int_{-1/2}^{1/2} \overline f(z) dz. \label{eq: vol}
\end{align}
It follows from the analysis of \cite{Lu2004} that the volume averages of $|\mathbf u|$ and $|T|$ are bounded uniformly in time, so spatiotemporal averages of time derivatives will vanish in what follows.

We shall focus on the volume averages of temperature, $\langle T \rangle$, and vertical convective heat flux, $\langle wT \rangle$, where $\mathbf u = (u,w)$. Bulk heat transport in RB convection is often characterized by a dimensionless quantity, the Nusselt number. We will see that $\langle T\rangle$ behaves like an inverse Nusselt number, while $\langle wT\rangle$ is a quite different quantity. To gain some understanding of the vertical structure of the flow, we will also consider horizontally averaged temperature profiles, $\overline T(z)$.

\subsection{Vertical heat fluxes}

Some physical understanding will be gained by examining the relative contributions of convection and conduction in the vertical direction. We denote the total vertical heat current by $J:=wT-\partial_zT$, where $wT$ is the convective part and $-\partial_zT$ is the conductive part. The average of $J$ over space and time is $\langle J \rangle = \langle wT\rangle$. Because the top and bottom boundary temperatures are equal, the conductive part of $J$ makes no net contribution to $\langle J\rangle$, reflecting the fact that the volume-averaged upward and downward conductive fluxes must equal one another at all times. Thus, any difference between the outward heat fluxes across the top and bottom boundaries is conveyed by $\langle wT\rangle$ alone, which may be thought of as the up-down asymmetry in vertical heat transport induced by fluid motion. Of course, convective and conductive processes are entwined in the details of the flow, and sustained fluid motion requires nonzero temperature gradients, so although vertical conduction vanishes in the mean, it is locally essential.

To quantitatively relate $\langle wT\rangle$ to the boundary fluxes, we first note that the mean outward heat flux across the top boundary is $-\overline T'_T$, and that across the bottom boundary is $\overline T'_B$, where the primes denote $d/dz$, and the subscripts denote evaluation at the top and bottom boundaries, respectively. These fluxes must always combine to equal the rate of heat production, which is normalized to unity, as is verified by integrating (\ref{eq: T}) to yield
\begin{equation}
\overline T'_B -\overline T'_T = 1. \label{eq: unity}
\end{equation}
In the conductive solution, both of the mean outward heat fluxes across the boundaries, $-\overline T'_T$ and $\overline T'_B$, have a value of 1/2, but a nonzero $\langle wT\rangle$ breaks this symmetry. Integrating $z\cdot(\ref{eq: T})$ reveals that $\langle wT\rangle$ is equal to half of the difference between the heat flowing out the top boundary and the heat flowing out the bottom one,
\begin{align}
\langle wT \rangle &= -\tfrac{1}{2} \big(\overline T'_T + \overline T'_B \big). \label{eq: zT}
\end{align}
Combining (\ref{eq: unity}) and (\ref{eq: zT}), we see that the (dimensionless) mean outward heat flux across the top boundary is $1/2+\langle wT\rangle$, while the outward flux across the bottom boundary is $1/2-\langle wT\rangle$. Since their sum is normalized to unity, these dimensionless fluxes also equal the fractions of the total produced heat that leave the domain via the top and bottom boundaries.

The mean vertical heat flux is bounded according to
\begin{equation}
0\le\langle wT \rangle \le 1/2, \label{eq: wT bound}
\end{equation}
as proven in \ref{sec: wT pf}. The lower bound, which corresponds to equal heat fluxes out of the top and bottom boundaries, is fulfilled only by the conductive solution. The upper bound corresponds to the maximally asymmetric case in which all heat flows out the top (that is, $-\overline T'_T=1$ and $\overline T'_B=0$). The nonnegativity of $\langle wT\rangle$ means that the onset of fluid motion can only increase heat flux across the top boundary and decrease heat flux across the bottom boundary. Physically, this is because fluid near the upper boundary has an adverse (negative) temperature gradient and drives the flow by sending relatively dense, cold fluid downward, while fluid near the lower boundary has a stabilizing (positive) temperature gradient and less readily sends cold fluid upward into the interior. Although $\langle wT \rangle$ is the quantity that arises naturally in integral relations, one may prefer to think in terms of the mean fraction of heat flowing out the top boundary, $1/2+\langle wT \rangle$, which lies between $1/2$ and $1$.

The horizontally averaged convective and conductive fluxes can be related by averaging (\ref{eq: T}) over $(0,A)\times (-1/2,z)$ and time to find \cite{Lu2004}
\begin{equation}
\overline{wT}(z)=\overline T'(z) - \overline T'_B + 1/2 + z,
\label{eq: slaving}
\end{equation}
which is why we need not examine $\overline{wT}(z)$ profiles in addition to $\overline T(z)$. Averaging $J$ over the horizontal and time, and applying (\ref{eq: slaving}), (\ref{eq: unity}), and (\ref{eq: zT}), yields $
\overline J = \langle wT \rangle + z$. That is, the mean heat flux across a horizontal surface increases linearly with height (because of the uniform volumetric heating) and is fully determined by the volume-integrated heat flux.

\subsection{Mean temperature}
\label{sec: T relations}

While $\langle wT\rangle$ conveys the difference in outward heat transport across the top and bottom boundaries, the mean fluid temperature, $\langle T\rangle$, conveys the relative amounts of convective and conductive transport responsible for carrying heat outward to the boundaries. To see this, we imagine the layer is divided along a plane where $\overline T(z)=\overline T_{max}$. A sensible value for the outward conductive heat flux is obtained by adding the magnitudes of (volume-averaged) upward conduction in the upper layer and downward conduction in the lower layer, which yields $2\overline T_{max}$. When $R$ becomes large, powerful fluid motion homogenizes the interior temperature, so that $\overline T_{max}\sim\langle T\rangle$. Thus, at a given large $R$, a lower value of $\langle T\rangle$ means that a higher fraction of the outward heat transport is achieved by convection, as opposed to conduction.

The dimensionless temperature is bounded according to
\begin{equation}
0<\langle T \rangle \le 1/12,
\end{equation}
as proven in \ref{sec: T pf}, though tighter $R$-dependent lower bounds exist \cite{Lu2004, Whitehead2011} and are stated in section \ref{sec: T results} below. The upper bound is saturated by the conductive solution, and $\langle T \rangle$ typically decreases with increasing $R$. This decrease may seem counter-intuitive until one recalls that the dimensionless $T$ is normalized by the rate of heat production.  If the rate of heat production is doubled, which doubles $R$, the dimensionful temperature of the layer will indeed increase, though it will not quite double. These diminishing returns are evinced by the decrease in dimensionless temperature.

\subsection{Power integrals}

Integrating $T\cdot(\ref{eq: T})$ and $\mathbf u \cdot (\ref{eq: u})$, respectively, yields \cite{Catton1974}
\begin{align}
\langle T \rangle &= \langle |\nabla T|^2 \rangle \label{eq: T PI} \\
R \langle wT \rangle &= \langle |\nabla \mathbf u|^2 \rangle. \label{eq: wT PI}
\end{align}
These relations correspond to the power integrals of RB convection \cite{Malkus1954a, Howard1963}. The RB power integrals serve as constraints for variational proofs of upper bounds on the Nusselt number \cite{Howard1972, Constantin1996}, as well as forming a basis for scaling arguments \cite{Grossmann2000}. As we argue in the next subsection, $\langle T \rangle$ behaves like an inverse Nusselt number, and indeed (\ref{eq: T PI}) and (\ref{eq: wT PI}) enable both scaling arguments (\emph{cf.}\ section \ref{sec: scalings}) and variational lower bounds \cite{Lu2004, Whitehead2011} for $\langle T \rangle$.

\subsection{Dimensionless numbers}
\label{sec: N}

The dimensionless quantities $\langle T\rangle$ and $\langle wT\rangle$ are natural choices for characterizing the mean vertical heat flux, in part because they arise in the power integrals (\ref{eq: T PI}) and (\ref{eq: wT PI}), yet neither of these quantities behaves like the Nusselt number of RB convection. The RB Nusselt number is bounded below and grows unboundedly with increasing Rayleigh number, whereas $\langle T\rangle$ and $\langle wT\rangle$ are bounded above and below. However, it is possible in the present case to define dimensionless quantities in place of $\langle T\rangle$ and $\langle wT\rangle$ that behave more like Nusselt numbers, and prior studies have done just that.

The Nusselt number of a developed flow is traditionally defined as the heat flux in a given direction (either across a surface or averaged over a volume), divided by the heat flux in the corresponding conductive (that is, static) solution. For the volume-averaged formulation of the Nusselt number in a plane layer, the convective and conductive fluxes are found by integrating $wT$ and $-\partial_zT$ in $z$. (We disregard any constant factors that may result from nondimensionalizing differently.) This yields the customary
\begin{equation}
N=\frac{\Delta \overline T_{dev}+\langle wT\rangle_{dev}}{\Delta \overline T_{cond}},
\label{eq: N}
\end{equation}
where the subscripts distinguish between the developed flow and the static conductive state, and $\Delta \overline T := \overline T_B - \overline T_T$.

This definition of Nusselt number serves well for standard RB convection, but it extends poorly to certain boundary conditions. When the boundaries have fixed fluxes instead of fixed temperatures, for instance, $N$ is identically unity. In light of this, some researchers have instead defined the Nusselt number as the volume-integrated heat flux in the developed flow, divided by the volume-integrated conductive flux in the developed flow \cite{Otero2002,Wittenberg2010,Johnston2009},
\begin{equation}
L=\frac{\Delta \overline T_{dev}+\langle wT\rangle_{dev}}{\Delta \overline T_{dev}}.
\label{eq: L}
\end{equation}
The quantity $L$ has useful behavior for a wider variety of boundary conditions than $N$, being unity in the conductive solution and typically growing unboundedly with Rayleigh number. However, $L=N$ only for fixed-temperature boundary conditions, so $L$ should not be confused with the traditional Nusselt number.

At large $R$, $\langle T\rangle$ is a sort of inverse $L$, for the following reason. Neither $L$ nor $N$ would be finite in our present case if defined using integrals over the entire layer because $\Delta \overline T$ would vanish. However, a useful volume-integrated conductive flux is recovered by considering only \emph{outward} flux. As discussed in section \ref{sec: T relations}, the outward conductive flux scales like $\langle T\rangle$ at large $R$. The total outward flux is unity, so an outward $L$ can be defined proportionally to $1/\langle T\rangle$. We simply focus on $\langle T\rangle$, but some intuition can be borrowed from the RB case by realizing that $1/\langle T\rangle$ behaves like a Nusselt number. On the other hand, it seems impossible to define a Nusselt number-like quantity in terms of $\langle wT\rangle$ alone.

Instead of using $\langle T\rangle$ and $\langle wT\rangle$, most previous studies of our configuration have characterized the bulk heat flow by two dimensionless quantities that are often called top and bottom Nusselt numbers (such studies are summarized in Figure 18 and Table 5 of \cite{Kulacki1985}). These quantities, which we denote by $L_T$ and $L_B$, are defined similarly to (\ref{eq: L}), but their numerators are surface fluxes instead of volume averages. The respective numerators of $L_T$ and $L_B$ are the outward heat fluxes across the top and bottom boundaries, $1/2 + \langle wT \rangle$ and $1/2 - \langle wT \rangle$, while the denominator of each is $\overline T_{max}$. Since $\overline T_{max}\sim \langle T \rangle$ for large $R$, we can say
\begin{align}
L_T &\sim \frac{1/2 + \langle wT \rangle}{\langle T \rangle} \label{eq: LT} \\
L_B &\sim \frac{1/2 - \langle wT \rangle}{\langle T \rangle}. \label{eq: LB}
\end{align}

In prior studies, both $L_T$ and $L_B$ have typically been fit to algebraic laws of the form $cR^\alpha$, as summarized in Table 5 of \cite{Kulacki1985}. However, our numerical results suggest that this is not an appropriate representation at large $R$, for the following reason. The fraction of heat flowing out the top, $1/2 + \langle wT \rangle$, is between 1/2 and 1, so the scaling of $L_T$ with increasing $R$ is simply that of $1/\langle T \rangle$.  The expectation that $L_B$ will have a different algebraic scaling implicitly assumes that $1/2 - \langle wT \rangle$ decays toward zero, and at moderate $R$ this seems to be the case.  In our large-$R$ numerical results, however, $\langle wT \rangle$ plateaus before reaching 1/2, so the scaling of $L_B$ becomes that of $1/\langle T \rangle$ as well. Thus, concentrating on the scalings of $L_T$ and $L_B$ becomes rather redundant at large $R$. Furthermore, the asymmetry between upward and downward heat flux is more clearly conveyed by $\langle wT \rangle$.  We therefore prefer to focus on $\langle T \rangle$ and $\langle wT \rangle$ instead of $L_T$ and $L_B$, though either pair of values may be approximately computed from the other according to (\ref{eq: LT}) and (\ref{eq: LB}).

\section{Quantitative results}
\label{sec: quant}

\subsection{Temperature profiles}

\begin{figure}
\includegraphics[scale=.75]{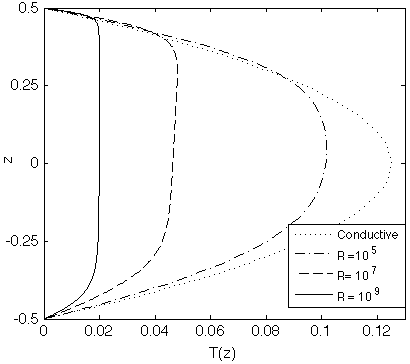}
\caption{Mean temperature profiles, $\overline T(z)$, for several $R$ with $\sigma=5$.}
\label{fig: profiles}
\end{figure}

Examples of mean vertical temperature profiles, $\overline T(z)$, are plotted in Fig.\ \ref{fig: profiles}. The onset of fluid motion is accompanied by an upward skewing of $\overline T(z)$, an increasingly isothermal interior, and an overall decrease in mean (dimensionless) temperature. By the time $R$ reaches $10^7$, the interior profile is very nearly linear, though a bit sub-isothermal (that is, stably stratified). As in RB convection, if the interior were completely isothermal, plumes would dissipate more slowly and overshoot further, and so restore the sub-isothermal conditions. Since the temperature profiles of Fig.\ \ref{fig: profiles} become visibly more asymmetric as $R$ increases, (\ref{eq: zT}) dictates that $\langle wT \rangle$ increases as well, at least for $\sigma=5$ and this range of $R$. Our computed values of $\langle wT\rangle$ indeed rise with $R$ in this regime, though not in all regimes.

\subsection{Convective heat flux}
\label{sec: wT results} 

\begin{figure}
\includegraphics[scale=.71]{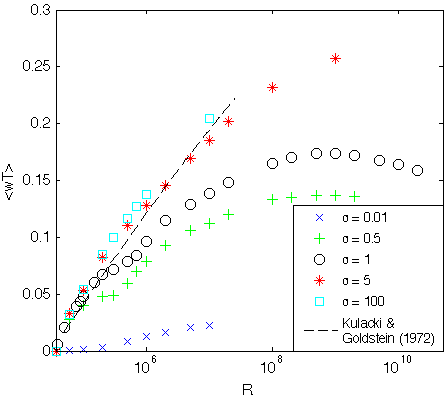}
\caption{Simulation results for mean vertical convective heat flux, $\langle wT\rangle$, beginning with a value of 0 at $R_L=37,325$.  (Adding 1/2 yields the fraction of produced heat flowing outward across the top boundary.)  Also shown is the fit given by Kulacki and Goldstein \cite{Kulacki1972} for their experimental data with $\sigma\approx6$.}
\label{fig: wT}
\end{figure}

Mean convective heat fluxes, $\langle wT \rangle$, are plotted in Fig.\ \ref{fig: wT} for various $R$ and $\sigma$, along with a fit proposed by Kulacki and Goldstein for their experimental data \cite{Kulacki1972}.  Over the $R$ range of their experiments, which employed an aqueous solution with $\sigma\approx6$, we find reasonable agreement between their fit and our $\sigma=5$ simulation results. This supports the claim \cite{Schmalzl2004} that the 2D system can be a good predictor of the 3D system's integral quantities for large enough $\sigma$.

While accurate over the range of $R$ in their experiments, the fit of Kulacki and Goldstein, like similar ones proposed for other moderate-$R$ data \cite{Jahn1974, Worner1997}, does not capture the behavior of $\langle wT \rangle$ that we observe at higher $R$. Such fits have typically been computed in terms of fits to $L_T$ and $L_B$ by
\begin{equation}
\langle wT \rangle + \frac{1}{2} = \frac{L_T}{L_T+L_B} \sim \frac{aR^\alpha}{aR^\alpha + bR^\beta}. 
\end{equation}
When applied to moderate-$R$ data, this ansatz will yield a larger growth rate with $R$ for $L_T$ than for $L_B$ (that is, $\alpha>\beta$), resulting in a fit for $\langle wT \rangle$ that asymptotes to 1/2 as $R\to\infty$. Such fits invariably exceed our computed values of $\langle wT \rangle$ at large $R$. This arrested growth of $\langle wT \rangle$ is visible in Fig.\ \ref{fig: wT} at the upper end of our $\sigma=5$ data, but it occurs at still lower $R$ for smaller $\sigma$. To explore the phenomenon further, we have carried the $\sigma=1$ and $\sigma=0.5$ simulations to $R=2\cdot10^{10}$ and $R=2\cdot10^9$, respectively.

Once $R$ exceeds roughly $10^9$ in our $\sigma=1$ and $\sigma=0.5$ simulations, $\langle wT \rangle$ not only stops growing but decreases with increasing $R$, as seen in Fig.\ \ref{fig: wT} . (This cannot be attributed to spatial under-resolution, which inflates $\langle wT \rangle$ due to under-resolved cold plumes descending farther before being warmed by thermal diffusion. This same effect inflates the Nusselt number in under-resolved simulations of 3D RB convection \cite{Stevens2010}.) The uncertainty in the data of Fig.\ \ref{fig: wT} is small, but it is still too large to determine whether $\langle wT\rangle$ starts decreasing first for $\sigma=1$ or for $\sigma=0.5$. The fact that $\langle wT\rangle$ does not asymptote to $1/2$ is also suggested by the superlinearity of $\log L_B$, plotted versus $\log R$, in the 3D simulation results of \cite{Worner1997}, which go up to $R=10^9$ for $\sigma=7$.

Competing physical mechanisms seem to be responsible for the initial increase and subsequent decrease of $\langle wT \rangle$ with increasing $R$. Initially, the cold down-flow from the unstable top boundary layer becomes stronger, as does the upward recirculation of warmer fluid, and this causes $\langle wT\rangle$ to grow. As noted in section \ref{sec: turb}, however, the increasingly strong down-flow begins to stir up cold fluid from the bottom boundary layer, which slows and subsequently reverses the growth of $\langle wT\rangle$. Still, since this cold up-flow is ultimately driven by the cold down-flow, it is surprising that the up-flow can strengthen with increasing $R$ faster than the down-flow does, which is necessary to explain the decrease in $\langle wT\rangle$ that we observe at large $R$.

The ultimate fate of $\langle wT \rangle$ as $R\to\infty$ remains uncertain. If $\langle wT\rangle\to0$ in the limit, meaning that heat ultimately flows equally out of both boundaries, this might be proven by a variational upper bound on $\langle wT \rangle$ that approaches zero in the limit. Such a result has eluded us, however. Whatever the fate of $\langle wT\rangle$, more experimental data would be useful. Our $R=2\cdot10^{10}$ simulation required two days on 256 BG/P processors for $\langle wT\rangle$ to converge. Such 2D numerics could be pushed to somewhat higher $R$, but perhaps not high enough, and the analogous computations in 3D would be extremely expensive. A physical experiment may be the best option.

\subsection{Temperature}
\label{sec: T results}

\begin{figure}
\includegraphics[scale=.7]{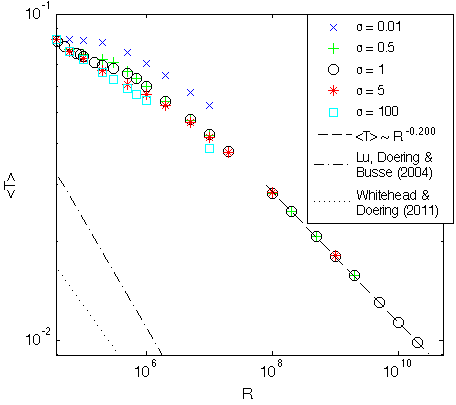}
\caption{Simulation results for mean dimensionless temperature, $\langle T\rangle$, beginning with a value of 1/12 at $R_L=37,325$. The algebraic fit (\ref{eq: fit}) to the last eight $\sigma=1$ data points is shown (dashed line), along with the lower bound of \cite{Lu2004} for arbitrary $\sigma$ (dash-dotted line) and the lower bound of \cite{Whitehead2011} for infinite $\sigma$ (dotted line).}
\label{fig: T}
\end{figure}

Mean temperatures, $\langle T \rangle$, are plotted in Fig.\ \ref{fig: T} for various $R$ and $\sigma$. Evidently, $\sigma$ has a weaker effect on $\langle T \rangle$ than on $\langle wT \rangle$. The $\sigma=1$ simulations were carried to high enough $R$ to reveal a nearly algebraic scaling of $\langle T \rangle$ with $R$, as reflected by a nearly straight line in the log-log plot of Fig.\ \ref{fig: T}. The data for Prandtl numbers of 0.5 and 5 fall nearly on the same line. The last eight $\sigma=1$ points are fit well by the law
\begin{equation}
\langle T \rangle = 1.13~R^{-0.200}.
\label{eq: fit}
\end{equation}
To three significant figures, the exponent of the fit is $-1/5$, which is one of the values predicted by our scaling arguments in the next subsection. (Performing the fit with one or two data points removed yields a range of exponents between $-0.196$ and $-0.204$.)

Also shown in Fig.\ \ref{fig: T} are the best known variational lower bounds on $\langle T \rangle$, which are consistent with our simulation results, if not tight. At leading order for large $R$, these bounds are $\langle T \rangle \ge 1.09 R^{-1/3}$ for all $\sigma$ \cite{Lu2004} and $\langle T \rangle \ge 0.419(R \log R)^{-1/4}$ in the infinite-$\sigma$ limit \cite{Whitehead2011}. The bounds were proven by Doering and collaborators using the integral constraints (\ref{eq: T PI}) and (\ref{eq: wT PI}) in application of the background method.  The infinite-$\sigma$ bound has a smaller prefactor, but it decays more slowly as $R\to\infty$, so it will ultimately be the tighter bound.  This is consistent with the next subsection's scaling arguments, which suggest that $\langle T \rangle$ decays more slowly in $R$ for larger $\sigma$.

All evidence indicates that, for stable, statistically steady solutions, $\langle T \rangle\to 0$ as $R\to\infty$. Certainly that is not true for all solutions: the conductive state has a mean temperature of 1/12 and solves the governing equations for all $R$, though it is unstable when $R$ is large.\footnote{Similarly, the Nusselt number of RB convection seems to grow unboundedly as $R\to\infty$ for stable, statistically steady states, while the conductive solution, for which $N=1$, reminds one that this need not be so for unstable solutions.}
If indeed $\langle T \rangle\to0$ for all stable solutions, proving as much may require new analytic machinery for restricting to stable solutions when constructing bounds.

\subsection{Scaling arguments}
\label{sec: scalings}

Over the past half century, several scaling arguments have been proposed to predict the dependence of Nusselt number on Rayleigh number in RB convection.  The approach of Grossmann and Lohse, put forth in \cite{Grossmann2000} and subsequent extensions \cite{Grossmann2001, Grossmann2004, Grossmann2011}, is the most systematic among them, predicting the existence of numerous scaling regimes in the $R$-$\sigma$ parameter plane.  As discussed in section \ref{sec: N}, our $\langle T \rangle$ is a kind of inverse Nusselt number, so we can apply the Grossmann-Lohse approach of \cite{Grossmann2000} in the search for scalings of $\langle T \rangle$. However, since the bulk heat flux in our problem is characterized by two numbers rather than one, the Grossman-Lohse arguments do not yield fully determined scalings; in general we cannot eliminate the unknown quantity $\langle wT \rangle$ from the scalings of $\langle T\rangle$. Obtaining scalings for both $\langle T\rangle$ and $\langle wT\rangle$ in terms of only $R$ and $\sigma$ would require a theory that addresses both boundary layers, whereas the Grossman-Lohse arguments adapted to the present problem involve only the top one. Nonetheless, we obtain some useful partial results.

As was done in \cite{Grossmann2000}, we assume that the interior flow is characterized by a single large-scale velocity, given in dimensionless terms by a Reynolds number, $Re$ (interpretations of which are discussed in \cite{Ahlers2009}), and we assume that the viscous boundary layer is a laminar one of Blasius type whose thickness scales as $\lambda_u\sim Re^{-1/2}$. (This and subsequent relations are dimensionless.) In our configuration, the temperature gradient of the top thermal boundary layer must remain of order unity since (\ref{eq: unity})-(\ref{eq: wT bound}) imply $1/2\le-\overline{T}'_T\le1$. Thus, the top thermal boundary layer's thickness must scale as $\lambda_T\sim \langle T\rangle$. The remainder of the argument consists of replacing the thermal dissipation, $\langle |\nabla T|^2\rangle$, in (\ref{eq: T PI}) and the viscous dissipation, $\langle |\nabla\mathbf u|^2\rangle$, in (\ref{eq: wT PI}) with scaling approximations. The approximations used depend on whether the dissipations are dominated by contributions from the boundary layers or the bulk, and on whether the viscous or thermal upper boundary layer is thicker. We omit the details of this procedure, as they are analogous to the RB case handled in \cite{Grossmann2000}, but the six intermediate relations that result are given in Table \ref{tab: intermediate}. From these six relations one obtains eight scalings of $\langle T\rangle$ and $Re$ in terms of $\sigma$ and $\mathcal R$, where $\mathcal R := R\langle wT\rangle$. The scalings of $\langle T\rangle$ are reported in Table \ref{tab: T}.

\begin{table}
\centering
\begin{tabular}{ccrl}
Dominant term & Thicker BL & \multicolumn{2}{c}{Scaling relation} \\
\hline
$\langle |\nabla \mathbf u|^2\rangle_{BL}$ & either & $R\langle wT\rangle \sim$ & \hspace{-8pt}$\sigma^2Re^{5/2}$ \\
$\langle |\nabla \mathbf u|^2\rangle_{bulk}$ & either & $R\langle wT\rangle\sim$ & \hspace{-8pt}$\sigma^2Re^3$ \\
$\langle |\nabla T|^2\rangle_{BL}$ & $T$ & $\langle T\rangle\sim$ & \hspace{-8pt}$\sigma^{-1/2}Re^{-1/2}$ \\
$\langle |\nabla T|^2\rangle_{BL}$ & $u$ & $\langle T\rangle\sim$ & \hspace{-8pt}$\sigma^{-1/3}Re^{-1/2}$ \\
$\langle |\nabla T|^2\rangle_{bulk}$ & $T$ & $\langle T\rangle\sim$ & \hspace{-8pt}$\sigma^{-1}Re^{-1}$ \\
$\langle |\nabla T|^2\rangle_{bulk}$ & $u$ & $\langle T\rangle\sim$ & \hspace{-8pt}$\sigma^{-1/2}Re^{-3/4}$ \\
\end{tabular}
\caption{Intermediate relations in the Grossmann-Lohse approach, depending on whether dissipations are dominated by the boundary layers or the bulk, and on whether the viscous ($u$) or thermal ($T$) upper boundary layer is thicker.}
\label{tab: intermediate}
\end{table}

\begin{table}
\centering
\begin{tabular}{cccp{22pt}l}
Dominant & Dominant & Thicker \\
$\langle |\nabla T|^2\rangle$ &
	$\langle |\nabla \mathbf u|^2\rangle$ & BL & \multicolumn{2}{c}{$\langle T \rangle$ Scaling} \\
\hline
BL & BL & $T$ & $\mathcal R^{-1/5}$& $\sigma^{-1/10}$ \\
BL & BL & $u$ & $\mathcal R^{-1/5}$& $\sigma^{1/15}$ \\
BL & bulk & $T$ & $\mathcal R^{-1/6}$& $\sigma^{-1/6}$  \\
BL & bulk & $u$ & $\mathcal R^{-1/6}$ \\
bulk & BL & $T$ & $\mathcal R^{-2/5}$& $\sigma^{-1/5}$ \\
bulk & BL & $u$ & $\mathcal R^{-3/10}$& $\sigma^{1/10}$ \\
bulk & bulk & $T$ & $\mathcal R^{-1/3}$& $\sigma^{-1/3}$ \\
bulk & bulk & $u$ & $\mathcal R^{-1/4}$ \\
\end{tabular}
\caption{Scalings of $\langle T \rangle$ with $\mathcal R := R\langle wT\rangle$ and $\sigma$ in eight different regimes, as predicted by arguments of Grossmann-Lohse type.}
\label{tab: T}
\end{table}

The simplest way to eliminate $\langle wT\rangle$ from the scaling laws of Table \ref{tab: T} is to assume that it remains $O(1)$, so that $\mathcal R\sim R$. We cannot justify this in general, but it is visibly true at the upper end of our $\sigma=1$ simulations, whose $\langle T \rangle$ scaling we would like to explain. In light of this, Table \ref{tab: T} suggests that our observed scaling of $\langle T \rangle \sim R^{-1/5}$ should occur when viscous and thermal dissipations are both dominated by their boundary layer contributions. In our high-$R$ simulations, $\langle |\nabla T|^2 \rangle$ is indeed boundary layer-dominated, but the bulk contribution to $\langle |\nabla \mathbf u|^2\rangle$ is several times larger than the boundary layer contribution.  In this regime, the scaling arguments instead predict $\langle T \rangle\sim R^{-1/6}$. This discrepancy is not caused by taking $\langle wT\rangle$ as constant; fitting $\langle T\rangle$ to a power of $\mathcal R$ rather than of $R$ merely changes the exponent from $0.200$ to $0.201$, and the fit is worse. One possible explanation of the discrepancy is that the apparent $R^{-1/5}$ scaling might be a mixture of the $R^{-1/6}$ regime with a sub-dominant $R^{-1/4}$ or $R^{-1/3}$ regime, which are the scalings expected when both dissipations are bulk-dominated.\footnote{Similarly, it is argued in \cite{Grossmann2000} that the apparent $N\sim R^{2/7}$ scaling sometimes seen in RB experiments \cite{Johnston2009, Castaing1989} may be interpreted as a superposition of $R^{1/3}$ and $R^{1/4}$ terms.}
 To determine whether this explanation is viable, one must locate the approximate boundaries in the $R$-$\sigma$ parameter plane that separate the various scaling regimes. This could be done using our data if one could unify the predicted scalings of $\langle T\rangle$ into a single function of $R$ and $\sigma$, in analogy to the way the scalings of \cite{Grossmann2000} are unified in \cite{Grossmann2001}. We have not attempted this because the lack of a theory for $\langle wT\rangle$ should be resolved before extending the scaling analysis in the manner of \cite{Grossmann2001, Grossmann2011}.

The scaling predictions of Table \ref{tab: T} are based on the assumption of a laminar viscous boundary layer, so they do not strictly apply when $R\to\infty$. Instead, we can employ a different scaling argument for this ultimate regime. When $R$ is very large, nearly all transport is achieved by convective turbulence, and this turbulence creates effective diffusivities different from the actual molecular parameters, $\kappa$ and $\nu$. Thus, we can argue that the mean dimensionful temperature should ultimately scale independently of the molecular parameters. The only scaling of dimensionless temperature with $R$ and $\sigma$ that satisfies this requirement is $\langle T\rangle\sim R^{-1/3}\sigma^{-1/3}$. This result suggests that $\langle T\rangle$ ultimately realizes the fastest rate of decay with $R$ that is permitted by the variational bound of \cite{Lu2004}.\footnote{In the analogous ultimate regime of RB convection, arguing that the dimensionful heat flux should scale independently of the molecular parameters leads to the Nusselt number scaling of $N\sim R^{1/2}\sigma^{1/2}$ \cite{Spiegel1971a}. This same scaling was derived by Kraichnan \cite{Kraichnan1962} using arguments based on shear layer turbulence, and it is implicit in treatments of convection zones in stars \cite{Spiegel1971b}.}

Our scaling arguments for $\langle T\rangle$ and the variational bounds on $\langle T\rangle$ computed in \cite{Lu2004,Whitehead2011} have both employed methods used previously to study the Nusselt number in RB convection. The success of these methods in the present case makes sense if we recall that $\langle T\rangle$ is roughly controlled by the top thermal boundary layer, and that the top half of our internally heated layer looks quite similar to the top half of a layer undergoing RB convection. To make quantitative the parallels between $1/\langle T\rangle$ and the RB Nusselt number, we can define a quantity $\hat R$ for the present problem that is more like the RB Rayleigh number than our $R$ is. The Rayleigh number is traditionally defined as $g\alpha d^3\Delta/\kappa\nu$, where $\Delta$ is the dimensionful temperature difference between the boundaries. We can define a similar quantity in the present case by using the mean temperature of the layer in place of $\Delta$. Doing so yields $\hat R := R\langle T\rangle$. The quantity $\hat R$ is useful for analysis, though it cannot replace $R$ as a control parameter because it requires knowledge of $\langle T\rangle$. In terms of $\hat R$, our predicted ultimate scaling is $1/\langle T\rangle\sim\hat R^{1/2}\sigma^{1/2}$, and the variational bound of \cite{Lu2004} is $1/\langle T\rangle\le c\hat R^{1/2}$. These expressions become identical to the corresponding RB results \cite{Spiegel1971a,Constantin1996} when we replace $1/\langle T\rangle$ with the Nusselt number and $\hat R$ with the traditional Rayleigh number. If $\langle wT\rangle$ is furthermore taken as constant (which is not always accurate), then the scalings of Table \ref{tab: T} become identical to the scalings computed for the RB problem by Grossmann and Lohse \cite{Grossmann2000}. All of these parallels reinforce the interpretation of $\langle T\rangle$ as an inverse Nusselt number.

\section{Conclusions}
\label{sec: con}

We have conducted direct numerical simulations of 2D internally heated convection for wide ranges of $\sigma$ and $R$, and we have presented both qualitative features and integral quantities. Qualitatively, the clearest differences from 2D RB convection are the frequent merging and genesis of downward-moving thermal plumes at moderate $R$ and the absence of upward-moving buoyant plumes. Quantitatively, we have focused on spatiotemporal averages of temperature, $\langle T \rangle$, and vertical convective heat flux, $\langle wT \rangle$, especially at large $R$. With $\sigma=1$, we obtained well-converged means for $R$ up to $2\cdot10^{10}$, higher than previously reported for direct simulation \cite{Jahn1974, Worner1997}. In this high-$R$ regime we observed unanticipated $R$-dependencies of both integral quantities.

Firstly, $\langle wT \rangle$ stops rising and begins decreasing with increasing $R$ in our high-$R$ simulations, meaning that the fraction of the total emergent heat that comes out the top is decreasing. In light of this, we have made the case that $\langle T \rangle$ and $\langle wT \rangle$ are more useful than the two quantities, called top and bottom Nusselt numbers, on which most previous studies have focused. Our reasons for this preference are that once $\langle wT \rangle$ stops increasing, both Nusselt numbers will scale with $R$ in the same way that $1/\langle T \rangle$ does, and that the Nusselt numbers do not convey the fraction of heat flowing out the top boundary as clearly as $\langle wT\rangle$ does. The ultimate fate of $\langle wT\rangle$ as $R\to\infty$ remains an open question.

Secondly, the dimensionless mean temperature scales as $\langle T \rangle\sim R^{-1/5}$ in our high-$R$ simulations. (This corresponds to the dimensionful mean temperature scaling with the heating rate like $H^{4/5}$.) We have presented Grossman-Lohse-type scaling arguments for $\langle T \rangle$ that predict the existence of up to eight different scalings in terms of $R\langle wT \rangle$ and $\sigma$, and which may explain the scaling we observed. We have also argued that $\langle T \rangle\sim R^{-1/3}\sigma^{-1/3}$ in the ultimate regime where $R\to\infty$ for order one $\sigma$. This is the same $R$-dependence as in the best known variational lower bound on $\langle T\rangle$ \cite{Lu2004}. We have argued that $\langle T\rangle$ is a kind of inverse Nusselt number, and the connection between $1/\langle T\rangle$ and the Nusselt number of RB convection was strengthened by re-expressing the scaling and bounding results for $\langle T\rangle$ in terms of $\hat R:=R\langle T\rangle$ and $\sigma$.

The scaling behaviors of integral quantities are of practical as well as theoretical interest. In engineering applications where sustained chemical or nuclear reactions are heating a fluid, the containing vessel must be able to carry heat away as fast as it is created, so that the temperature will not rise indefinitely. The vessel must also be able to withstand localized spikes in temperature due to the inhomogeneity of temperature in the fluid. Such spikes in boundary temperature cannot occur with the isothermal boundaries we have imposed, but in real-world cases where the boundaries are good but imperfect conductors, we can reasonably expect that the maximum instantaneous temperatures to which the boundaries are subjected will scale with $R$ roughly as $\langle T\rangle$ does. Knowing $\langle wT \rangle$, on the other hand, tells us the relative fractions of the produced heat that the top and bottom boundaries must be capable of carrying away. When the boundaries are imperfect conductors but still identical to one another, the larger heat flux out the top will keep the top boundary hotter than the bottom one, which works against upward heat transport and should decrease the difference between upward and downward heat fluxes. Therefore, among all configurations with identical top and bottom boundaries, our results should provide upper bounds on the fraction of heat flowing out the top boundary, an interesting result to try proving analytically.

Several open questions remain, and they invite a multifaceted attack. Further 2D simulations may help identify other scaling regimes of $\langle T \rangle$, testing the scaling arguments we have presented. Meanwhile, a better physical understanding of how the bottom thermal boundary layer interacts with the bulk could lead to predictions for the parameter-dependence of $\langle wT \rangle$ and more complete predictions for the scaling of $\langle T\rangle$.  But we expect that the most promising avenue will be physical experiment. Obtaining well-converged integral quantities in 3D direct simulations may be prohibitively computationally intensive for the largest $R$ that we have simulated in 2D. However, laboratory experiments could access such Rayleigh numbers, and they may even illuminate the asymptotic fate of $\langle wT \rangle$.

\section*{Acknowledgments}

D.G.\ thanks Charles Doering for guidance on several aspects of this work, David Keyes for general advice and support, and Paul Fischer and Aleksandr Obabko for much help running nek5000. We thank Francis Kulacki and Antonello Provenzale for stimulating conversations about this work, Luis Fernandez for a helpful observation, and Cody Ranaldo for assistance with image preparation. D.G.\ was supported by the NSF under the project EMSW21 - RTG: Numerical Mathematics for Scientific Computing (Award No.\ DMS-0602235). Some of our computing resources were provided by the New York Center for Computational Sciences at Stony Brook University/Brookhaven National Laboratory, which is supported by the DOE (Contract No.\ DE-AC02-98CH10886) and by the State of New York. E.A.S.\ is grateful to the members of the Applied Mathematics Laboratory of the Courant Institute at NYU for their unfailingly warm hospitality during the preparation of this letter.

\appendix

\section{Integral bounds}

\subsection{Proof that $T>0$ on the interior}
\label{sec: min}

Let $\Omega$ be an open, bounded domain in any number of spatial dimensions.  Assume $T$ remains smooth on $\Omega$ for all $t\ge0$, that it vanishes on the boundary of $\Omega$, and that $T>0$ everywhere on the interior of $\Omega$ at the initial time, $t=0$. To prove the result by contradiction, suppose that $T<0$ on an interior point at some positive time. There then must exist a minimum time, $t_0$, at which $T$ crosses zero and at least one interior point, $x_0$, at which this occurs. That is, $T\ge0$ everywhere when $t\le t_0$, but $T<0$ everywhere on some neighborhood of $x_0$ for some time interval, $(t_0,t_1)$. However, $T$ attains a spatial minimum at $(x_0,t_0)$, so $\mathbf u \cdot \nabla T=0$ and $\nabla^2 T \ge 0$, from which (\ref{eq: T}) implies $\partial_t T(x_0,t_0) > 0$.  Hence, $T$ will be positive at $x_0$ as soon as $t$ exceeds $t_0$, a contradiction.

\subsection{Proof that $0<\langle T \rangle\le1/12$}
\label{sec: T pf}

The nonnegativity of $\langle T \rangle$ follows easily from (\ref{eq: T PI}).  If we also assume regularity of $T$, meaning smooth temperature solutions exist for all time, then the result of \ref{sec: min} applies to give strict positivity of $\langle T\rangle$. (With much more effort, the background method analysis of \cite{Lu2004} gives a better lower bound for all but the smallest values of $R$.)

To obtain the upper bound, we integrate $z^2\cdot(\ref{eq: T})$ to find
\begin{equation}
\langle T \rangle = \tfrac{1}{12}-\langle z w T \rangle. \label{eq: z^2T}
\end{equation}
It thus suffices to show that $\langle zwT \rangle$ is nonnegative.  Integrating the continuity equation over the horizontal and using the side boundary conditions gives $\overline w(z) \equiv 0$, from which it follows that $\langle z w T \rangle=\langle z w \theta \rangle$, where $\theta$ is the deviation of $T$ from the conductive profile, $\widetilde T$. The PDE governing $\theta$ is
\begin{equation*}
\partial_t \theta + \mathbf u \cdot \nabla \theta - z w = \Delta \theta,
\label{eq: T pert}
\end{equation*}
and integrating this equation against $\theta$ gives $\langle z w \theta \rangle=\langle |\nabla \theta|^2\rangle$, which is nonnegative. Thus, $\langle T \rangle \le 1/12$.

\subsection{Proof that $0\le\langle wT \rangle\le1/2$}
\label{sec: wT pf}

The nonnegativity of $\langle wT \rangle$ for positive $R$ follows directly from (\ref{eq: wT PI}). To obtain the upper bound, we assume regularity of $T$.  Then, positivity of $T$ on the interior (\emph{c.f.} \ref{sec: min}) implies $\overline{T}'_B \ge 0$ and $\overline{T}'_T\le 0$.  The latter inequality in conjunction with Equation~(\ref{eq: unity}) requires that $\overline{T}'_T \ge -1$ as well.  Applied to Equation~(\ref{eq: zT}), the lower bounds on $\overline{T}'_B$ and $\overline{T}'_T$ together give $\langle wT \rangle \le 1/2$.

\section{Computational methods}
\label{sec: comp}

The nek5000 code \cite{nek} was run with second-order variable time stepping with a target Courant number of 0.5 on up to 256 parallel processors. Spatiotemporal averages were deemed converged at a time, $\tau$, when the cumulative averages $\langle T \rangle$ and $\langle wT \rangle$ each differed by less than 0.2\% from their values at $\tau/2$.  Similarly, spatial meshes were deemed converged when increasing the polynomial order of each element by 2 produced a change of less than than 0.5\% in the three spatiotemporal averages.

The element meshes used were tensor products, with $n_x$ equal-width elements in the horizontal and $n_z$ elements in the vertical. The vertical element spacing followed Gauss-Lobatto-Chebyshev points, so element heights scale as $1/n_z^2$ near the boundaries and as $1/n_z$ near the center. This helps avoid under-resolving the boundary layers, which would create much larger errors than under-resolving the interior. We fixed $n_x/n_z=2A/3$, based on resolution studies performed with $R=10^6$ and $\sigma=1$. The finest mesh used was for the simulation with $R=2\cdot10^{10}$ and $\sigma=1$, for which $n_z=96$ with order-6 elements, yielding about 14 points in the top thermal boundary layer and more in the bottom one.

For sufficiently large $A$, either insulating or periodic side boundaries would suffice for volume averages to approximate those of an infinite domain, but when $R$ is large enough for the flow to be unsteady, periodic domains were found to converge faster as $A\to\infty$. (For steady flows, however, averages may converge faster on domains with insulating sides because any integer number of convection rolls is possible, while periodic domains require an even number of rolls, which can force the flow farther from its preferred horizontal scale.) We performed several aspect ratio studies, which together suggested that $A=3$ with periodic sides approximates the infinite domain sufficiently when $R\gtrsim10^7$ at moderate-to-large $\sigma$. For instance, spatiotemporal averages changed by less than 1\% when the aspect ratio was increased from 3 to 9 with $\sigma=1$ and $R=10^7$. Similarly, we found $A=9$ sufficient at all smaller $R$, so long as the flow was aperiodic.

\section{Descriptions of ancillary videos}
\label{sec: videos}

The videos described below are available as ancillary files on arXiv.

\begin{enumerate}

\item (Supplement to Fig.\ \ref{fig: fast}) Evolution of the fluid speed (top) and temperature (bottom) fields over the fast phase of an oscillation with $R=65,000$, $\sigma=1$, and $A=12$. The merging and genesis of cold thermal plumes (or equivalently, of rolls) is evident. The speed scale begins at zero (black) and saturates at 15 (white). The temperature scale begins at zero (blue) and saturates at 0.12 (red). The dimensionless time of the video is 3, whereas the slow phase of an oscillation has a period on the order of 100.

\item (Supplement to Fig.\ \ref{fig: plumes}) Typical evolution of the temperature field with $R=10^8$ and $\sigma=5$, and over a dimensionless time of 0.015. The video is cropped to an aspect ratio of 7 from a wider domain. The temperature scale begins at zero (blue) and saturates at 0.036 (red).

\item (Supplement to Fig.\ \ref{fig: plumes}) Typical evolution of the temperature field with $R=10^{10}$ and $\sigma=5$, and over a dimensionless time of 0.002. The video is cropped to an aspect ratio of 7 from a wider domain. The temperature scale begins at zero (blue) and saturates at 0.0135 (red). Some numerical artifacts are visible near the top boundary; $\langle T\rangle$ and $\langle wT\rangle$ values from this simulation were not included in Fig.\ \ref{fig: wT} or \ref{fig: T}.

\item (Supplement to Fig.\ \ref{fig: low Pr}) Typical evolution of the fluid speed (top) and temperature (bottom) fields with $R=2\cdot10^5$ and $\sigma=0.01$, and over a dimensionless time of 3.92. The video is cropped to an aspect ratio of 7 from a wider domain. The speed scale begins at zero (black) and saturates at 7 (white). The temperature scale begins at zero (blue) and saturates at 0.125 (red), which is the maximum temperature of the static state.

\item (Supplement to Fig.\ \ref{fig: low Pr}) Typical evolution of the fluid speed (top) and temperature (bottom) fields with $R=10^7$ and $\sigma=0.01$, and over a dimensionless time of 0.297. The video is cropped to an aspect ratio of 7 from a wider domain. The speed scale begins at zero (black) and saturates at 125 (white). The temperature scale begins at zero (blue) and saturates at 0.09 (red).

\item (Highest-$R$ simulation) Typical evolution of the temperature field with $R={2\cdot10^{10}}$, $\sigma=1$, and $A=3$, and over a dimensionless time of $10^{-4}$. This is the largest $R$ for which we obtained well converged $\langle T\rangle$ and $\langle wT\rangle$ values (see Fig.\ \ref{fig: wT} and \ref{fig: T}). The temperature scale begins at zero (blue) and saturates at 0.012 (red). In addition to cold plumes descending from the top boundary layer, cold eddies are evidently being shed by the bottom boundary layer.

\bibliographystyle{model1a-num-names}
\bibliography{library}

\end{enumerate}

\end{document}